# Contextuality-by-Default Description of Bell Tests: Contextuality as the Rule and Not as an Exception

Marian Kupczynski

Département d'informatique et d'ingénierie, Université du Québec en Outaouais, Case Postale1250, Succursale Hull, Gatineau, QC J8X 3X7, Canada; marian.kupczynski@uqo.ca

**Abstract:** Contextuality and entanglement are valuable resources for quantum computing and quantum information. Bell inequalities are used to certify entanglement; thus, it is important to understand why and how they are violated. Quantum mechanics and behavioural sciences teach us that random variables 'measuring' the same content (the answer to the same Yes or No question) may vary, if 'measured' jointly with other random variables. Alice's and Bob's raw data confirm Einsteinian non-signaling, but setting dependent experimental protocols are used to create samples of coupled pairs of distant ±1 outcomes and to estimate correlations.Marginal expectations, estimated using these final samples, depend on distant settings. Therefore,a system of random variables 'measured' in Bell tests is inconsistently connected and it should be analyzed using a Contextuality-by-Default approach, what is done for the first time in this paper.The violation of Bell inequalities and inconsistent connectedness may be explained using a contextual locally causal probabilistic model in which setting dependent variables describing measuring instruments are correctly incorporated. We prove that this model does not restrict experimenters` freedom of choice which is a prerequisite of science. Contextuality seems to be the rule and not an exception; thus, it should be carefully tested.

**Keywords:** Bell inequalities; counterfactual definiteness and noncontextuality; quantum nonlocality; Einsteinian non-signaling;entanglement; local realism; measurement independence; Kochen–Specker contextuality; Bohr complementarity

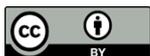

## 1. Introduction

In classical physics, we describe a world, as we perceive it, in terms of non-contextual and contextual properties. The dimensions and shape of a marble table are non-contextual. They are believed to exist before they are measured and they do not depend on when and how they are measured. Measurements, with a good approximation, are non-invasive and their outcomes give the information about these pre-existing properties. On other hand, a color of the chameleon is contextual, because it depends on where it is observed.

In quantum physics, measurements are invasive, and their outcomes are created in interaction of a physical system with measuring instruments, in well defined experimental context. If an experimental context is changed, quantum probabilistic description is modified. Therefore, we say that quantum observables are contextual. There exist incompatible experimental contexts in which incompatible quantum observables are measured.

In Bell tests we have four incompatible random experiments. Local realistic and stochastic hidden variables models failed to explain outcomes of Bell tests, because they described 'entangled photons' as pairs of socks or as pairs of fair dice.

In this article, we discuss in detail non-contextual and contextual properties, along with the random variables used to describe them. Some estimated marginal expectations in Bell tests depend on distant settings, and the corresponding random variables are inconsistently connected. Therefore, we use Contextuality-by-Default (CbD) to study them, which is the main subject of this paper.

A population is a fundamental concept in mathematical statistics. It may be a set of physical systems, objects, animals, or people (whose properties, behaviour and opinions at a given moment of time) we want to investigate.It is also an infinite set of outcomes, which might have been obtained in subsequent repetitions of some random experiment performed in an unchanged experimental context.

The information about statistical populations is inferred from properties of finite samples. This information is reliable only; if we have at our disposal simple random samples drawn from the population we are investigating.

Various physical observables and properties are coded using continuous or discrete random variables, and 'measured' values of these random variables may be displayed in various spreadsheets. We say that a set of properties and random variablesrepresenting themare non-contextual, if they can be 'measured' in any order and a studied population may be described by a joint probability distribution of these random variables.

In some sense, non-contextual properties, characterizing members of a population, are believed to exist independently of the fact of being measured. This is why we may use joint probability distributions of random variables to describe populations, for which not all of these random variables can be measured jointly. In statistical physics, we even use with success joint probability distributions of impossible to measure positions, linear momenta, and energies of invisible molecules, in order to describe thermodynamics of materials.

Experiments in quantum physics also involve invisible physical systems and we observe the macroscopic effects of their interactions with measuring instruments or environment:traces left by charged particles in various ionization chambers, clicks on detectors, etc. Clicks on detectors are interpreted as values of some physical observables 'measured' in the experiment. A statistical scatter of these values obtained in a series of 'identical repeated measurements' performed on 'identically prepared physical systems' is compared with quantum predictions.

In classical physics, measuring instruments 'register' (with limited precision and possible errors) pre-existing values of non-contextual observable.If measurements are performed on different members of a population, a scatter of outcomes is only due to the fact, that a studied population is a mixed statistical ensemble. A simple example of such an ensemble is a box which contains equal number of red and black 'identical items' from which we draw with replacement one "item" at the time.Classical filters are devices which select objects having different pre-existing properties

There is a fundamental difference between classical and quantum measurements, thus as Bohr [1–3] insisted we should rather talk about quantum experiments and quantum phenomena.

Quantum theory teaches us that the outcomes of measurements are created in an interaction of a physical system with a measuring instrument in a well-defined experimental context [4]. Incompatible quantum instruments/filters 'create' complementary physical properties which may not be measured jointly, and their values may not be assigned 'to the same physical system' at the same time [5].

Let us consider, for example, a monochromatic laser beam linearly polarized in a direction **n**. Its intensity is measured by number of clicks on a single-photon detector. Since the intensity of the beam is not significantly changed, if we pass this beam by another polarization filter directed in the direction **n**, we conclude that all 'photons' are polarized in the same direction **n**. However, if we pass this beam by another polarization filter directed in the direction **m** ≠ **n**, the intensity of the beam diminishes according to the Malus law and all remaining 'photons' are polarized now in the direction **m**.It may be easily checked that they are no longer polarized in the direction **n**.

It is difficult to construct consistent 'mental images' of 'quantum objects' because atomic phenomena are characterized by: "*the impossibility of any sharp separation between the behaviour of atomic objects and the interaction with the measuring instruments which serve to define the conditions under which the phenomena appear*" (Bohr ([1], v. 2, p. 40–41).

In a recent paper, Andrei Khrennikov resumed these fundamental features of atomic phenomena in two principles [6]:

*Bohr-contextuality: The output of any quantum observable is indivisibly composed of the contributions of the system and the measurement apparatus.*

*Bohr-complementarity: There exist incompatible observables (complementary experimental contexts).*

If (A, B, C) are only pair-wise measurable observables, then to 'measure' A and B on a physical system we must use different experimental instruments/contexts, than when we 'measure' A and C. Thus Bohr-contextuality allows qualitative understanding of Kochen–Specker-contextuality [7] (as we define it):

*KS-contextuality: A measurement of an observable does not need to yield the same value independently of what other measurements may be made simultaneously*[7–16].

*KS-contextuality* is not limited to quantum mechanics. In cognitive sciences answers to Yes-or-NO questions given by an individual depend on which other questions are asked at the same time and on a whole experimental context. Therefore Dzhafarov and Kujala pointed out, that random variables describing outcomes of these experiments should be labelled not only by content but also by a context of an experiment. In their approach called *Contextuality-by-Default* (CbD) [17], the same questions are represented by different random variables depending on what other questions are asked at the same time.

Their approach applies also to experiments in physics and in other domains of science [17–23] and even it may be generalised. In physics we register time series of results subdivided often into successive runs of the same experiment. It is not sure that all reproducible properties of this time series may be explained completely using a probability distribution of a single random variable. As we pointed out several years ago [24–26], it has to be tested and not taken for granted. Therefore, to be on safe grounds, each experimental run may be described by a different random variable, and one has to verify whether they are identically distributed or not.

Similarly, in behavioural and cognitive sciences, outcomes of experiments performed on different samples drawn from the same population might be described by different random variables. Only by comparing gathered data we can decide whether these, a priori different, random variables may be considered as the same or not (in different words whether our finite samples are *simple random samples* drawn from the same population).

Differences between finite samples drawn from the same population have nothing to do with *KS-contextuality*, which is an intimate context dependent relation between studied random variables.

In mathematical statistics, multivariate random variables and joint probability distributions are only used to describe random experiments or population surveys, in which each trial/individual is described not by one, but by several data items. In this case we say that these data items are 'measured' values of commeasurable random variables.

Einstein believed that quantum pure ensembles are in fact mixed statistical ensembles of physical systems [27,28], which may be described by joint probability distributions of non-contextual hidden variables (NCHV). In such probabilistic models, pairwise marginal expectations must obey some noncontextuality inequalities (NCI) which are violated by quantum expectations and by experimental data.

The violation of NCI, in Bell tests [29–36] is often interpreted as the violation of *local realism*. In our opinion one should rather talk about *naïve realism* = noncontextuality or *non-invasive measurability*. This violation, as we explain in this paper, is only a manifesta-tion of Bohr- and KS-*contextuality* and has nothing to do with the locality or non-locality of nature. Outcomes of experiments in quantum physics and in cognitive sciences are not predetermined, before the experiments are done, and they depend strongly on experimental protocols, and on experimental contexts.

Many authors tried to explain the true meaning of Bell-type inequalities and of their violations [37–89]. They arrived, often independently to the conclusion, that Bell inequalities are only the necessary and sufficient conditions for the existence of a counterfactual joint probability distribution describing outcomes of incompatible random experiments. They pointed out that, if hidden variables describing measuring instruments are correctly incorporated in the probabilistic models, Bell inequalities may not be derived. Titles of the cited papers are self- explanatory, but the discussion of them is out of the scope of this article.

Nevertheless, speculations about quantum nonlocality and quantum magic are still quite frequent in blogs, books, and scientific papers. Such unfounded speculations are not only motivated by the violation of Bell inequalities but also by incorrect interpretations of quantum mechanics, as it was clearly explained in several recent papers [53–55,66–70,89–97].

NCI are also violated in experiments in social, behavioural and cognitive sciences [20,98–101], this is why when discussing the results of Bell tests, we will only talk about *contextuality* and not about *non-locality*. We also agree with Peres that *unperformed experiments have no results* [76].

The paper is organized as follows.

In Section 2, we explain what we mean by *contextuality of a set of random variables*, and we present general n-cyclic NCI [101].

In Section 3, we recall the main assumptions and definitions used in CbD [17–19, 23].

In Section 4, we discuss experimental protocols used in Bell tests. We explain observed context dependence of marginal distributions (*inconsistent connectedness* of random variables) and we derive modified CHSH inequality allowing studying contextuality of these random variables more in detail.

In Section 5, we explain *inconsistent connectedness* in Bell Tests, using a contextual locally causal model, in which setting dependent variables describing measuring instruments are correctly introduced.

In Section 6, we reject the incorrect and often repeated claim that setting dependence of variables in a probabilistic model restricts *experimenters' freedom of choice*.

In Section 7, we explain why contextuality should be the rule in spin polarization correlation experiments and we propose new experimental tests.

## 2. Contextuality and Non-Contextuality

If physical systems/individuals have properties/opinions, at a given moment of time, which do not depend on, whether they are 'measured'/asked–for, then we may describe various statistical populations of these systems/individuals by a joint probability distribution of non-contextual random variables.

Contextual properties/opinions of systems/individuals do not exist before being 'measured'/asked-for in a particular experimental context.Therefore, if we have a set of random variables which are not all commeasurable usually there is a deep reason for it and the results of experiments may not be explained by assuming the existence of a *counterfactual joint probability distribution* of all these variables.

Let us consider a set of random variables $X=\{X_0…X_{n-1}\}$ which may be measured on members of a statistical population. We propose a general and experimentally testable definition of *contextuality*.

> If not all variables in a set X are commeasurable, then a set X is called contextual, if one may reject a statistical hypothesis that a studied population is described by a joint probability distribution of all these variables. Otherwise, the set is called non-contextual.

*A set X of dichotomous random variables, taking values ±1, is contextual, if and only if various NCI are significantly violated.*

Of particular importance are NCI satisfied by cyclic expectation values of pairs of random variables $<X_0 X_1>$, $<X_1 X_2>$…$<X_{n-1}X_0>$:

$$< X_0 X_1 > + < X_1 X_2 > +…+ < X_{n-2}X_{n-1} > - < X_{n-1}X_0 > \leq n-2 \qquad (1)$$

The inequality (1) follows immediately from a simple arithmetic inequality: $x_0 x_1 + x_1 x_2 +…+ x_{n-2}x_{n-1} - x_{n-1}x_0 \leq n-2$, which is always satisfied by $x_i = \pm 1$. For $n=3$ the inequality (1) is in fact one of Boole [102] or Suppes–Zanotti–Legett–Garg (SZLG) inequalities [103,104]. For $n=4$ we obtain Clauser–Horn–Shimony–Holt (CHSH) inequality [105] and for $n=5$ we obtain Klyachko–Can–Binicioglu–Shumovsky (KCBS) inequality [106].

In a similar way one proves all generalized n-cyclic NCI for $n \geq 3$:

$$\sum_{i=0}^{n-1} \gamma_i < X_i X_{i+1} > \leq n-2 \qquad (2)$$

where $X_n = X_0$, $\gamma_i \in \{-1,1\}$ and the number of $\gamma_i = -1$ is odd [19,101].

## 3. Contextuality-by-Default Approach

In CbD approach, random variables measuring the same content in different contexts are *stochastically unrelated*, and they are labelled by contexts, in which they are measured. Dzhafarov and Kujala explain it clearly in several articles [19–23]. In this paper we use a simplified notation, similar to that of Araujo et al. [101], which suits better our discussion of Bell tests.

Let us consider an n-cycle scenario of binary random variables $X=\{X_0 …, X_{n-1}\}$ such that only all successive pairs $\{X_i, X_{i+1}\}$ are commeasurable and $X_n = X_0$.

Since each pair of random variables defines a different experimental context, thus we have a new system containing 2n binary random variables $X'=\{X_0, X'_0…X_{n-1}, X'_{n-1}\}$. We have still *n* pairwise measurable expectation values $<X_i X'_{i+1}>$, but now random variables $X_i \neq X'_i$ are stochastically unrelated (our system is not cyclic) and we may not derive inequalities (2).

If marginal expectation values measured in different contexts violate marginal selectivity/parameter independence $<X_i>_m \neq <X'_i>_m$, we say that these random variables, representing the same content in different contexts, are *inconsistently connected* (NCC), otherwise they are *consistently connected* (CC). *Inconsistent connectedness* is the first indication that a system exhibits KS-*contextuality*, but in CbD one wants to obtain more detailed answers to two questions [23]:

*"A: For any two random variables, sharing content, how different are they when taken in isolation from their contexts?*

*B: Can these differences be preserved when all pairs of content-sharing variables are taken within their contexts (i.e., taking into account their joint distributions with other random variables in their contexts)?"*

This is why Dzhafarov and Kujala generalised NCI for NCC systems. Any set containing stochastically related and stochastically unrelated random variables can always be coupled (imposed a counterfactual joint probability distribution upon) [17–19]. If such probability distribution is imposed, expectations $<X_i X'_i>$ are defined and we have a new 2n-cyclic system/scenario for which one may derive immediately NCI similar to (1) and (2):

$$< X_0 X'_0 > + < X'_0 X_1 > + < X_1 X'_1 > +…+ < X_{n-1}X'_{n-1} > - < X'_{n-1}X_0 > \leq 2n-2 \qquad (3)$$

$$\sum_{i=0}^{n-1} \gamma_i < X'_i X_{i+1} > + \sum_{i=0}^{n-1} < X_i X'_i > \leq 2n-2 \qquad (4)$$

Since the random variables $X_i$ and $X'_i$ correspond to the same content in different contexts, they should be as similar as possible, what imposes constraints on $<X_iX'_i>$. If such constraints are imposed, then *a counterfactual joint probability distribution* of 2n variables, consistent with experimental data, does not always exist and the violation of the inequalities (3) and (4) allows us to study a degree of contextually of the system $X'$.

In CbD we impose the *maximal coupling* on each pair of random variables $\{X_i, X'_i\}$ replacing $<X_iX'_i>$ by its maximal value consistent with observed marginal expectations $<X_i>_m$ and $<X'_i>_m$. As it was proven in [19] (Lemma 3):

*Jointly distributed ±1-valued random variables A and B with given expectations $<A>$; $<B>$; $<AB>$ exist if and only if:*

$$|\langle A\rangle+\langle B\rangle|-1 \leq \langle AB\rangle \leq 1-|\langle A\rangle-\langle B\rangle| \qquad (5)$$

After replacing $<X_iX'_i>$ in (4) by their maximal values, evaluated using the equation (5), we obtain (<u>in our notation</u>) Dzhafarov-Kujala NCI:

$$\sum_{i=0}^{n-1}\gamma_i <X'_iX_{i+1}> + \sum_{i=0}^{n-1}[1-|\langle X_i\rangle_m-\langle X'_i\rangle_m|] \leq 2n-2 \qquad (6)$$

By rearranging terms in (6) and replacing $<X_i>_m$ by $<X_i>$ we obtain a simpler and more transparent NCI:

$$S_n = \sum_{i=0}^{n-1}\gamma_i <X'_iX_{i+1}> - \sum_{i=0}^{n-1}|\langle X_i\rangle-\langle X'_i\rangle| \leq n-2 \qquad (7)$$

where $n\geq 3$, $\gamma_i \in \{-1,1\}$ and the number of $\gamma_i = -1$ is odd.

If all $<X_i> = <X'_i>$, the maximal coupling becomes the identical coupling $<X_iX'_i>=1$ and we recover inequalities (2) after replacing $X'_i$ by $X_i$. Thus for CC systems and $n=4$, $S_4 = S$ and (7) is the well-known CHSH inequality.

If the maximal coupling exists, then according to CbD the system $X'$ has *maximally non-contextual description* and is called *non-contextual*. However one should not forget that the *significant inconsistent connectedness* is already a manifestation of *KS-contextuality*, <u>as we define it</u>, and that the violation of inequalities (7) by experimental data is an <u>additional</u> important measure of *contextuality* of $X'$.

Kujala, Dzhafarov, and Larsson studied the violation of the inequality (7) for KCBS system [106], using experimental data of Lapkiewicz et al.[107]. They assessed the significance of the violation of (7) using Bonferroni confidence intervals. This method can be easily generalised for any values of $n\geq 3$. If $I_\alpha(y) = [l_\alpha, u_\alpha]$ is an estimated $(1-\alpha)100\%$ confidence interval for an unknown population parameter $y$, then there is $(1-\alpha)100\%$ chance, $(\Pr(y \in I_\alpha(y)) = 1-\alpha)$, that the value of this parameter is included in $I_\alpha(y)$.

If we define: [a, b] + [c, d] = [a+ c, b+ d] and - [a, b] = + [−b, −a], then a conservative $I_\alpha(S_n)$ may be written as follows:

$$I_\alpha(S_n) = \sum_{i=0}^{n-1}\gamma_i I_{\frac{\alpha}{2n}}(<X'_iX_{i+1}>) - \sum_{i=0}^{n-1}I_{\frac{\alpha}{2n}}(|\langle X_i\rangle-\langle X'_i\rangle|) \qquad (8)$$

If the lower bound of $I_\alpha(S_n)$ is greater than $n-2$, then with $(1-\alpha)100\%$ confidence, we may conclude that $X'$ is *strongly* contextual (it does not allow *maximally non-contextual description*). If an upper bound of $I_\alpha(S_n)$ is smaller than $n-2$, then we may conclude with $(1-\alpha)100\%$ confidence that $X'$ allows *maximally non-contextual description*.

It is often believed that the CbD approach is not of much use for Bell tests, because according to Einsteinian non-signaling principle random variables measured by Alice and Bob should not depend on what is measured in a distant location.

In the next section, we show that Einsteinian non-signaling is not violated in Bell tests. Nevertheless, random variables describing samples, extracted from raw data and used to estimate correlations, exhibit *inconsistent connectedness* and they should be ana-lyzed using CbD approach. The violation of Bell-type inequalities is due to the *contextuality* of quantum observables and to context dependent protocols needed to establish coupling between outcomes of distant measurements. It has nothing to do with *quantum nonlocality* [53–55,66–70,89–97].

## 4. Contextuality-by-Default Description of Bell Tests

There are essential differences between impossible to implement experimental protocol of EPRB experiment [108,109] and experimental protocols used in SPCE to demonstrate the violation of CHSH inequality [30, 31].

In the EPRB thought experiment we have a steady flow of twin-electron or twin-photon pairs. Alice and Bob, working in distant laboratories, measure spin projections using 4 pairs of settings (i, j) = (x, y), (x, y'), (x', y), and (x', y'), which define contexts of four incompatible experiments. Outcomes for each twin-photon pair are coded by values of random variables ($A_i$, $B_j$), where $A_i$ = ±1 and $B_j$ = ±1. There are strict correlations and anti-correlations for some settings and marginal expectations $<A_i>$ = $<B_j>$ = 0 as predicted by QM. There are no losses of pairs, and all expectations $<A_i B_j>$ may be unambiguously estimated using experimental data [67].

In a typical spin polarization correlation experiment (SPCE), two correlated signals ("twin-photon beams") are sent from a source to Alice's and Bob's polarization beam splitters and detectors, which we call: PBS-detector modules. Pair emissions are governed by some stochastic process not described by QM. A click on a detector is interpreted as the detection of a photon with "spin up" or "spin down" in a particular direction. There are black counts, laser intensity drifts, photon registration time delays, etc. Each detected click, coded as 1 or −1, has its time tag and raw data are samples of two stochastically unrelated time-series. Several steps are needed to extract from raw data final samples, allowing to establish a coupling between distant outcomes and to estimate correlations between them. A much more detailed discussion may be found in [67,84,85,110]. Here, we enumerate only 3 steps of the experimental protocol for a fixed setting (x, y):

1. Raw time-tagged data are two samples: $S_A(x, y)$ = {($a_k$, $t_k$) | k=1…$n_x$} and $S_B(x, y)$ = {($b_m$, $t'_m$) | j=1…$n_y$}, with $a_k$ = ±1 and $b_m$ = ±1.
2. Using fixed synchronized time-windows of width W and keeping only windows, in which there is no click at all or a click on one of Alice's or/and Bob's detectors, new samples are created: $S_A(x, y, W)$ = {$a_s$ | s=1,…$N_x$} and , $S_B(x, y, W)$ = {$b_t$ | t = 1…$N_y$}, with $a_s$ = 0, ±1 and $b_t$ = 0, ±1.
3. Now by keeping only synchronized time-windows, in which both Alice and Bob observed a click on one of their detectors, a new sample of paired outcomes is obtained: $S'_{AB}(x, y, W)$ = {($a_r$, $b_r$) | r=1,…$N_{xy}$}, with $a_r$ = ±1 and $b_r$ = ±1.

In fact we have a large family of samples labelled by W and a corresponding family of random variables [67], but one chooses the optimal value of W which maximizes the number of coincidences.

If samples constructed in the step 2 are used, then $<A|x, y, W>$ ≈ $<A|x, y', W>$ and $<B|x, y, W>$ ≈ $<B|x', y, W>$, thus Einsteinian non-signaling (parametric independence) is not violated in SPCE.

In step 2, the random variables A and B are equal to 0 or ±1. To test CHSH inequality we have to estimate expectations $<A'B'|x, y, W>$, $<A' B'|x, y', W>$, $<A' B'|x', y, W>$ and $<A'B'|x', y', W'>$ using samples constructed in step 3. Now A' and B' are new random variables equal to ±1.

Adenier and Khrennikov [110] and De Raedt, Jin, and Michielsen [84,85] analyzed the raw data of Weihs et al. [31] and discovered that marginal expectations $<A'|i, j>$ and $<B'|i, j>$ depended on distant settings. This apparent violation of parameter independence/non-signaling, could not be explained by the violation of a fair sampling assumption, and was in conflict with quantum predictions.

Similar anomalies were discovered by Adenier and Khrennikov [111] and by Bednorz [112] in Hensen et al. data [33]. Liang, using the work of Lin et al. [113] and of Zhang et al. [114] analyzed the data from [115] and reported, at FQMT2017, that the probability (*p*-value) of observing some data points, under the assumption of non-signaling, was smaller than $3.17 \times 10^{-55}$. The results were derived assuming that measurement settings were randomly chosen, but it turned out that this assumption was not respected in the experiment of [115]. A detailed discussion of these results was published in a recent paper [116].

Moreover, in this experiment, as in many other Bell tests [65,117], it was not checked carefully enough that trials are independent and identically distributed. We demonstrated with Hans de Raedt [118], that without such verification the standard statistical inference is not reliable. A detailed discussion of experimental protocols and possible loopholes in Bell tests may be found in Larsson [119].

Apparent violation of Einsteinian non-signaling reported in [110–114,116] is only the effect of context dependent experimental protocols required to establish correlations between clicks on distant detectors. It is also a manifestation of Bohr-*contextuality* and may be explained in a locally causal way using context dependent variables describing PBS-detector modules [67,70].

Random variables A' and B' measuring, in different contexts, the same content (presence of a click on one of Alice's and Bob's detectors) are *inconsistently connected*, thus CbD is the appropriate approach to study more in detail *contextuality* of this NCC system.

In CbD the random variables A' and B' are labelled by corresponding contexts/settings (i, j). To simplify the notation, we replace (x, y) by (1, 1) etc.

We have now a system X' = $\{A_{11}, A_{12}, A_{21}, A_{22}, B_{11}, B_{12}, B_{21}, B_{22}\}$ of 8 binary random variables (describing 4 samples obtained in step 3 of the experimental protocol), which is inconsistently connected ($<A_{11}> \neq <A_{12}>, <A_{21}> \neq <A_{22}>, <B_{11}> \neq <B_{21}>, <B_{12}> \neq <B_{22}>$). After introducing maximal couplings, as it was explained in the preceding section, the system X' is transformed into 8-cyclic system.

Therefore, instead of CHSH inequality:

$$S = <A_1 B_1> + <A_1 B_2> + <A_2 B_1> - <A_2 B_2> \leq 2 \tag{9}$$

we obtain in CbD a new inequality for $S_4$:

$$S_4 = <A_{11} B_{11}> + <A_{12} B_{12}> + <A_{21} B_{21}> - <A_{22} B_{22}> - D_4 \leq 2 \tag{10}$$

where $D_4$ is the contribution from 4 counterfactual maximal couplings:

$$D_4 = |<A_{11}> - <A_{12}>| + |<B_{11}> - <B_{21}>| + |<A_{21}> - <A_{22}>| + |<B_{12}> - <B_{22}>| \tag{11}$$

The violation of inequality (10) allows assessing more precisely a degree of *contextuality* of X'. It may be done using conservative confidence intervals (8) for $S_4$.

$$I_\alpha(S_4) = I_{\frac{\alpha}{8}}(<A_{11} B_{11}>) + \ldots I_{\frac{\alpha}{8}}(<A_{21} B_{21}>) - I_{\frac{\alpha}{8}}(<A_{22} B_{22}>) - I_{\frac{\alpha}{2}}(D_4) \tag{12}$$

In the Bell tests, we have 4 random experiments performed in incompatible experimental settings. Each of these experiments may be described by its own Kolmogorov probability space and the only constraint, which may be derived, without assuming *non-contextuality* or by imposing maximal couplings, is $|S| \leq 4$.

In the next section, we present a contextual probabilistic model able to explain in a locally causal way data obtained in step 3 of the experimental protocol discussed above.

## 5. Contextual Locally Causal Probabilistic Model

The inconsistently connected random variables describing the experimental data may neither be explained using quantum mechanical model for EPRB nor by local real-istic hidden variable models, because in these models the parameter independence is strictly obeyed.

As demonstrated in [67,70], the apparent violation of non-signaling and inconsistent connectedness may be explained by a contextual probabilistic model in which setting dependent variables describing measuring devices are correctly introduced.

- Photonic signals arriving to PBS-detector modules are described by variables $(\lambda_1, \lambda_2) \in \Lambda_1 \times \Lambda_2$ and p $(\lambda_1, \lambda_2)$.
- In a setting (x, y), Alice's and Bob's instruments, at the moment of measurement, are described by variables $(\lambda_x, \lambda_y) \in \Lambda_x \times \Lambda_y$ and probability distributions $p_x(\lambda_x)$ and $p_y(\lambda_y)$.
- Outcomes 0, ±1 are the values of functions $A_x(\lambda_1, \lambda_x)$ and $B_y(\lambda_2, \lambda_y) = 0, \pm 1$.

Expectation values of inconsistently connected random variables A'=$A_{xy}$ and B'=$B_{xy}$, describing data obtained in step 3 of the experimental protocol are given by:

$$E(A_{xy}B_{xy}) = E(A_xB_y \mid A_xB_y \neq 0) = \sum_{\lambda \in \Lambda'_{xy}} A_x(\lambda_1, \lambda_x) B_y(\lambda_2, \lambda_y) p_{xy}(\lambda) \quad (13)$$

$$E(A_{xy}) = E(A_x \mid A_xB_y \neq 0) = \sum_{\lambda \in \Lambda'_{xy}} A_x(\lambda_1, \lambda_x) p_{xy}(\lambda) \quad (14)$$

$$E(B_{xy}) = E(B_y \mid A_xB_y \neq 0) = \sum_{\lambda \in \Lambda'_{xy}} B_y(\lambda_2, \lambda_y) p_{xy}(\lambda) \quad (15)$$

where $p_{xy}(\lambda) = p_x(\lambda_x) p_y(\lambda_y) p(\lambda_1, \lambda_2)$, $\Lambda_{xy} = \Lambda_1 \times \Lambda_2 \times \Lambda_x \times \Lambda_y$ and

$$\Lambda'_{xy} = \{\lambda \in \Lambda_{xy} \mid A_x(\lambda_1, \lambda_x) \neq 0, B_y(\lambda_2, \lambda_y) \neq 0\} \quad (16)$$

For each setting (x, y), data obtained in step 2 of the experimental protocol are described, by random variables $A_x$ and $A_y$ obeying a joint probability distribution $p_{xy}(\lambda)$ on a specific probability space $\Lambda_{xy}$. Since $\Lambda_{xy} \cap \Lambda_{xy'} \cap \Lambda_{x'y} \cap \Lambda_{x'y'} = \emptyset$, CHSH and other Bell inequalities may not be derived.

It is incorrectly believed, that the dependence of hidden variables on settings in a probabilistic model restricts *experimenters' freedom of choice* or *measurement independence*. In the next section, we explain why it is not true.

## 6. Contextuality Does Not Restrict *Experimenters' Freedom of Choice*

Despite the fact that there is no agreement as to why Bell-type inequalities are violated, most of authors agree that the proof of these inequalities is based on 3 main assumptions [120]:

1. Realism
2. Locality
3. Freedom of choice, measurement independence or no-conspiracy

In a recent paper Blasiak et al. [121] conclude that the violation of the free choice assumption is an important resource in Bell experiments. It is surprising, because as Bell said [120,122,123]:

> *"It has been assumed that the settings of instruments are in some sense free variables—say at the whim of the experimenters—or in any case not determined in the overlap of the backward light cones. Indeed without such freedom I would not know how to formulate any idea of local causality, even the modest human one."*

This point of view is probably shared by the majority of physicists [36].

Fortunately, the assumption 3, as it is used to prove Bell inequalities, has nothing to do with *experimenters' freedom of choice*. The misunderstanding consists on an incorrect interpretation of conditional probabilities and Bayes Theorem [66,124].

*Measurement independence* is often defined as: *measurement settings can be chosen independently of any underlying variables describing the system*. This definition is rephrased using conditional probabilities [121,125]:

$$p(x, y, \lambda) = p(x, y) p(\lambda), p(x\, y|\lambda) = p(x, y), p(\lambda|x, y) = p(\lambda) \tag{13}$$

Equation (17) resumes correctly a mathematical content of the assumption 3, but $p(\lambda|x, y) = p(\lambda)$ means only that variables $\lambda$, describing signals, do not depend on a choice of settings, and that there exists a joint probability distribution of these variables on a unique probability space $\Lambda$, which may be used to describe the outcomes of four incompatible random experiments. Therefore, Equation (17) is only *noncontextuality* assumption, which is closely related to *realism* (the assumption 1), as we defined it in the introduction.

*By measurement independence or freedom of choice* we understand something more general: *measurement settings can be chosen independently of any underlying variables describing the experiment and its outcomes*.

In our contextual model (13−16):

$$p(x, y, \lambda) = p_{xy}(\lambda)p(x, y) = p(\lambda), p(\lambda|x, y) = p_{xy}(\lambda), \mathbf{p(x\, y|\lambda) = p(x, y, \lambda)/p(\lambda) = 1}. \tag{14}$$

The equation $\mathbf{p(x\, y|\lambda)=1}$ means only: *if an 'event {$\lambda=(\lambda_1, \lambda_{2x}, \lambda_y)$}' occurred, thus, the settings (x, y) were used*. It does not mean that {λ} had any *causal influence* on a choice of the settings [66,124]. It is visualized in Figure 1.

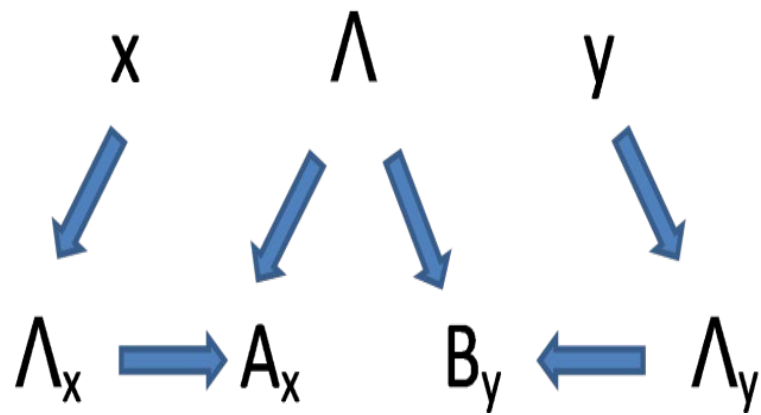

**Figure 1.** Experimenters choose freely their settings. A choice of settings is not only a choice of labels (x,y), but also it is a choice of spaces $\Lambda_x$ and $\Lambda_y$ describing the instruments in these settings. The outcomes $A_x$ and $A_y$ are created in a locally causal way. They are determined by the variables describing instruments and the variables $\Lambda$ describing photonic signals at the moment of their interaction.

In our model $p(\lambda|x, y) \neq p(\lambda)$ and $p(x, y|\lambda) \neq p(x, y)$ but *experimenters' freedom of choice*, which is a prerequisite of science, is fully respected. A much more detailed discussion of conditional probabilities and of equations (17) may be found in [66].

## 7. Discussion

The violation of various Bell-type inequalities clearly demonstrated that the values ±1, denoting clicks on detectors are not pre-existing properties of incoming signals, as it was assumed in local realistic hidden variable models. Clicks are macroscopically magnified effects of an interaction of correlated signals with PBS-detector modules.

The violation of Bell-type inequalities has nothing to do with *magical quantum nonlocality*. It is a manifestation of Bohr- contextuality. As Karl Svozil wrote in a recent paper [126]:

> "One could understand Bohr and Bell also by their insistence that the value definite properties (characterizing its physical state) of the object become "amalgamated" with (properties of) the measurement apparatus, so that an observation signals the combined information both of the object as well as of the measurement apparatus."

A particularity of Bell tests is that we want to estimate correlations between outcomes of <u>distant</u> experiments. We <u>have to create a coupling</u> of these outcomes (a procedure which is rarely unambiguous) in order to obtain a sample described by a generalized joint probability distribution of two random variables. As we explained in [5] a standard joint probability distribution does not exist for the outcomes of distant experiments. In Bell tests we have four incompatible random experiments and only rarely such experiments may be described using a generalized joint probability distribution on the unique probability space.

The inconsistent connectedness reported in Bell tests [84,85,110–114,116] is neither consistent with quantum description of EPRB experiments nor with local realistic hidden variable models. It may easily be explained using a contextual probabilistic model in which setting dependent variables describing measuring instruments are correctly incorporated [66,67,70].

As we explained in the preceding section the so-called *measurement independence* assumption is simply *noncontextuality* assumption. Therefore, its violation in our contextual model does not restrict *experimenters' freedom of choice*.

The important message for the quantum information community is that contrary to what was claimed [121], the true resource in Bell experiments is neither *nonlocality* nor *freedom of choice* but *contextuality*.

Inconsistent connectedness seems to challenge a quantum mechanical description of Bell Tests. Similarly, as Calude et al. reported [127], large sequences of random bits, generated from the detection of photons, were incapable of passing some randomness tests like Borel normality. Martinez et al. [128] explained, that the unwanted correlations are introduced by the APD detectors due to after pulsing and dead time. Because of these and other biases quantum random number generators (QRNGs) actually perform rather poorly in tests of randomness as compared to classical pseudo-random number generators (PRNGs).

In SPCE the context dependent step 3 of experimental protocol does not depend on how signals are correlated at the source. Moreover Bohr-*contextuality* should not depend on how settings are chosen. Therefore, one could expect that *inconsistent connectedness* and a violation of inequalities [10] may be observed not only for particular (angles), choices of settings, and not only for beams of "entangled twin–photon pairs", but also for different photonic signals. In order to gather larger samples, settings do not need to be changed randomly, when photons are in flight. They could be fixed in advance and kept the same during a long experimental run. One may even check whether the results depend on how the settings are chosen and changed. We do not believe that it will make a difference.

Such tests focused on studying the *inconsistent connectedness* and other anomalies in SPCE are needed to answer the following question:

> *What is more important cause of the violation of Bell-type inequalities: a particular entanglement of incoming signals and a choice of particular settings or Bohr- and KS-contextuality and context dependent experimental protocols?*

In our opinion *contextuality* in SPCE should be the rule and not an exception. Our conjecture seems to be confirmed by Iannuzzi, Francini, Messi and Moricciani [129], who recently reported the violation of Bell inequalities in the experiments with independent sources of polarized photons:

> *"We present a Bell-type polarization experiment using two independent sources of polarized optical photons and detecting the temporal coincidence of pairs of uncorrelated photons which have never been entangled in the apparatus. The outcome of the experiment gives evidence of violation of the Bell-like inequalities".*

They used different experimental protocols, than the protocols discussed in this article. Nevertheless, their results seem to prove that, in their experiment, the inequalities are violated mainly because of *contextuality* and not because of *entanglement*.

The violation of NCI in Bell scenario is often called *nonlocality* instead of *contextuality*. *Contextuality* has nothing to do with *nonlocality*, by which one usually understands spooky influences, or instantaneous transfer of information between distant experimental set-ups [54,70]. Such influences would have been necessary, if perfect fair dices had produced perfectly correlated outcomes in distant locations. Fortunately, such experiments do not exist. Our model (13–16) explains imperfect correlation in Bell tests in a locally causal way, without requiring any interactions between distant signals and instruments.

In this article, we concentrated on (probabilistic) contextuality, and on CbD approach.Since contextuality is an important resource for quantum computing [130–132], thus its different aspects and measures have been studied intensively using several different approaches. Let us mention here the sheaf-theoretic approach of Abramsky–Brandenburger [133], the graph approach of Cabello–Severini–Winter [134], the hypergraph approach of Acin et al.[135], and the operational approach of Spekkens [136].

**Funding:** This research received no external funding.

**Institutional Review Board Statement:** Not applicable.

**Informed Consent Statement:** Not applicable.

**Data Availability Statement:** No experimental data are analyzed in this paper.

**Acknowledgments:** I want to thank anonymous referees of this article for several valuable suggestions. I am also indebted to Andrei Khrennikov for his hospitality extended to me during several conferences on quantum foundations in Växjö.

**Conflicts of Interest:** The author declares no conflict of interest.

**References**

1. Bohr, N. *The Philosophical Writings of Niels Bohr*; Ox Bow Press: Woodbridge, UK, 1987.
2. Bell, J.S. On the problem of hidden variables in quantum theory. *Rev. Mod. Phys.* **1966**, *38*, 450.
3. Plotnitsky, A. *Niels Bohr and Complementarity: An Introduction*; Springer: New York, NY, USA, 2012.
4. Nieuwenhuizen, T.M.; Kupczynski, M. The contextuality loophole is fatal for derivation of Bell inequalities: Reply to a comment by I. Schmelzer. *Found. Phys.* **2017**, *47*, 316–319, doi:10.1007/s10701-017-0062-y.
5. Kupczynski, M. Bell inequalities, experimental protocols and contextuality. *Found. Phys.* **2015**, *45*, 735–753.
6. Khrennikov, A. Can there be given any meaning to contextuality without incompatibility? *Int. J. Theor. Phys.* **2020**, doi:10.1007/s10773-020-04666-z.
7. Kochen, S.; Specker, E.P. The problem of hidden variables in quantum mechanics. *J. Math. Mech.* **1967**, *17*, 59–87.
8. Greenberger, D.M.; Horne, M.A.; Shimony, A.; Zeilinger, A. Bell's theorem without inequalities. *Am. J. Phys.* **1990**, *58*, 1131–1143.
9. Mermin, N.D. Hidden variables and the two theorems of John Bell. *Rev. Mod. Phys.* **1993**, *65*, 803–815.
10. Peres, A. *Quantum Theory: Concepts and Methods;* Kluwer: Dordrecht, The Netherlands, 1995.
11. Cabello, A.; Estebaranz, J.; Garcìa-Alcaine, G. Bell- Kochen-Specker theorem: A proof with 18 vectors. *Phys. Lett. A* **1996**, *212*, 183–187.
12. Simon, C.; Brukner, C.; Zeilinger, A. Hidden-variable theorems for real experiments. *Phys. Rev. Lett.* **2001**, *86*, 4427–4430.
13. Cabello, A. Experimentally testable state-independent quantum contextuality. *Phys. Rev. Lett.* **2008**, *101*, 210401.
14. Yu, S.; Oh, C.H. State-independent proof of Kochen-Specker theorem with 13 rays. *Phys. Rev. Lett.* **2012**, *108*, 030402.
15. Cabello, A. Simple explanation of the quantum violation of a fundamental inequality. *Phys. Rev. Lett.* **2013**, *110*, 060402.
16. Winter, A. What does an experimental test of quantum contextuality prove or disprove? *J. Phys. A: Math. Theor.* **2014**, *47*, 424031.
17. Dzhafarov, E.N.; Kujala, J.V. Contextuality is about identity of random variables. *Phys. Scr.* **2014**, *T163*, 014009.

18. Dzhafarov, E.N.; Kujala, J.V.; Larsson, J.-Å. Contextuality in three types of quantum-mechanical systems. *Found. Phys.* **2015**, *7*, 762-782.
19. Kujala, J.V.; Dzhafarov, E.N.; Larsson, J-Å Necessary and sufficient conditions for extended noncontextuality in a broad class of quantum mechanical systems. *Phys. Rev. Lett.* **2015**, *115*, 150401.
20. Cervantes, V.H.; Dzhafarov, E.N. Snow Queen is evil and beautiful: Experimental evidence for probabilistic contextuality in human choices. *Decision* **2018**, *5*, 193-204.
21. Dzhafarov, E.N. On joint distributions, counterfactual values, and hidden variables in understanding contextuality. *Philos. Trans. R. Soc. A* **2019**, *377*, 20190144.
22. Kujala, J.V.; Dzhafarov, E.N. Measures of contextuality and noncontextuality. *Philos. Trans. R. Soc. A* **2019**, *377*, 20190149. (available as arXiv:1903.07170.).
23. Dzhafarov, E.N. Contents, contexts, and basics of contextuality. In *From Electrons to Elephants and Elections: Saga of Content and Context*; The Frontiers Collection; Wuppuluri, S., Stewart, I., Eds.; Springer: Berlin/Heidelberg, Germany, 2021. In press.
24. Kupczynski, M. Tests for the purity of the initial ensemble of states in scattering experiments. *Lett. Nuovo Cim.* **1974**, *11*, 121–124.
25. Kupczynski, M. On some new tests of completeness of quantum mechanics. *Phys. Lett. A* **1986**, *116*, 417–419.
26. Kupczynski, M. Is quantum theory predictably complete? *Phys. Scr.* **2009**, *T135*, 014005. Doi:10.1088/0031-8949/2009/T135/014005
27. Kupczynski, M. Time series, stochastic processes and completeness of quantum theory. *AIP. Conf. Proc.* **2011**, *1327*, 394-400.
28. Schilpp, P.A. (Ed.) *Albert Einstein: Philosopher–Scientist*; Harper and Row: New York, NY, USA, 1949.
29. Einstein, A. Physics and reality. *J. Frankl. Inst.* **1936**, *221*, 349.
30. Aspect, A.; Grangier, P.; Roger, G. Experimental test of Bell's inequalities using time-varying analyzers. *Phys. Rev. Lett.* **1982**, *49*, 1804–1807.
31. Weihs, G.; Jennewein, T.; Simon, C.; Weinfurther, H.; Zeilinger, A. Violation of Bell's inequality under strict Einstein locality conditions. *Phys. Rev. Lett*. **1998**, *81*, 5039–5043.
32. Christensen, B.G.; McCusker, K.T.; Altepeter, J.B.; Calkins, B.; Lim, C.C.W.; Gisin, N.; Kwiat, P.G. Detection-loophole-free test of quantum nonlocality, and applications. *Phys. Rev. Lett.* **2013**, *111*, 130406.
33. Hensen, B.; Bernien, H.; Dréau, A.E.; Reiserer, A.; Kalb, N.; Blok, M.S.; Ruitenberg, J.; Vermeulen, R.F.L.; Schouten, R.N.; Abellán, C.; et al. Loophole free Bell inequality violation using electron spins separated by 1.3 kilometres. *Nature* **2015**, *526*, 682–686.
34. Hensen, B.; Bernien, H.;Dreau, A.E.; Reiserer,A.; Kalb, N.; Blok, M.S.; Ruitenberg, J.: Vermeulen, R.F.L.; Schouten, R.N.; Abellan C; et al. Significant-loophole-free test of Bell's theorem with entangled photons. *Phys. Rev. Lett.* **2015**, *115*, 250401.
35. Shalm, L.K.; Meyer-Scott, E.; Christensen, B.G.; Bierhorst, P.; Wayne, M.A.; Stevens, M.J.; Gerrits, T.; Glancy, S., Hamel, D.R.; Allman, M.S.; et al. Strong loophole-free test of local realism. *Phys. Rev. Lett.* **2015**, *115*, 250402.
36. Abellán, C.; Acín, A.; Alarcón, A.; Alibart, O.; Andersen, C.K.; Andreoli, F.; Beckert, A.; Beduini, F.A.; Bendersky, A.; Bentivegna, M.; et al. The BIG Bell Test collaboration challenging local realism with human choices. *Nature* **2018**, *557*, 212–216, doi:10.1038/s41586-018-0085-3.
37. Accardi, L. Topics in quantum probability. *Phys. Rep.* **1981**, *77*, 169-192.
38. Accardi, L. Some loopholes to save quantum nonlocality. *AIP. Conf. Proc.* **2005**, *750*, 1–19.
39. Aerts, D. A possible explanation for the probabilities of quantum mechanics. *J. Math. Phys.* **1986**, *27*, 202–210.
40. Czachor, M. On some class of random variables leading to violation of the Bell inequality. *Phys. Lett. A* **1988**, *129*, 291. new version (2017) arXiv:1710.06126 [quant-ph].
41. Czachor, M. Arithmetic loophole in Bell's Theorem: Overlooked threat to entangled-state quantum cryptography. *Acta Phys. Polon. A* **2021**, *139*, 70–83, doi:10.12693/APhysPolA.139.70.
42. Fine, A. Hidden variables, joint probability and the Bell inequalities. *Phys. Rev. Lett.* **1982**, *48*, 291–295.
43. Fine, A. Joint distributions, quantum correlations, and commuting observables. *J. Math. Phys.* **1982**, *23*, 1306–1310.
44. Hess, K.; Philipp, W. Bell's theorem: Critique of proofs with and without inequalities. *AIP Conf. Proc.* **2005**, *750*, 150–157.
45. Hess, K.; De Raedt, H.; Michielsen, K. Hidden assumptions in the derivation of the theorem of Bell. *Phys. Scr.* **2012**, *T151*, 014002.
46. Hess, K.; Michielsen, K.; De Raedt, H. From Boole to Leggett-Garg: Epistemology of Bell-type Inequalities. *Adv. Math. Phys.* **2016**, *2016*, 4623040, doi:10.1155/2016/4623040.
47. Jaynes, E.T. Clearing up mysteries—The original goal. In *Maximum Entropy and Bayesian Methods*; Skilling, J., Ed.; Kluwer: Dordrecht, The Netherlands, 1989; Volume 36, pp. 1–27, doi:10.1007/978-94-015-7860-8_1.
48. Khrennikov, A. *Interpretations of Probability*; VSP Int. Tokyo, Japan; Sc. Publishers: Utrecht, The Netherlands, 1999.
49. Khrennikov, A. Bell-boole inequality: Nonlocality or probabilistic incompatibility of random variables? *Entropy* **2008**, *10*, 19–32.
50. Khrennikov, A. *Contextual Approach to Quantum Formalism*; Springer: Dordrecht, The Netherlands, 2009.
51. Khrennikov, A. *Ubiquitous Quantum Structure*; Springer: Berlin, Germany, 2010.
52. Khrennikov, A. CHSH inequality: Quantum probabilities as classical conditional probabilities. *Found. Phys.* **2015**, *45*, 711.
53. Khrennikov, A. Get rid of nonlocality from quantum physics. *Entropy* **2019**, *21*, 806.
54. Khrennikov, A. Two Faced Janus of Quantum Nonlocality. *Entropy* **2020**, *22*, 303, doi:10.3390/e22030303.
55. Khrennikov, A. Quantum versus classical entanglement: Eliminating the issue of quantum nonlocality. *Found. Phys* **2020**, *50*, 1762–1780, doi:10.1007/s10701-020-00319-7.
56. Kupczynski. M. *New test of Completeness of Quantum Mechanics;* International Atomic Energy Agency: Trieste, Italy, 1984.

57. Kupczynski, M. Pitovsky model and complementarity. *Phys. Lett. A* **1987**, *121*, 51–53.
58. Kupczynski, M. Bertrand's paradox and Bell's inequalities. *Phys. Lett. A* **1987**, *121*, 205–207.
59. Kupczynski, M. Entanglement and Bell inequalities. *J. Russ. Laser Res*. **2005**, *26*, 514-523.
60. Kupczynski, M. Seventy years of the EPR paradox. *AIP Conf. Proc*. **2006**, *861*, 516–523.
61. Kupczynski, M. EPR paradox, locality and completeness of quantum. *AIP Conf. Proc*. **2007**, *962*, 274–285.
62. Kupczynski, M. Entanglement and quantum nonlocality demystified. *AIP Conf. Proc*. **2012**, *1508*, 253–264.
63. Kupczynski, M. Causality and local determinism versus quantum nonlocality. *J. Phys. Conf. Ser*. 2014, 504 012015, doi:10.1088/1742-6596/504/1/012015.
64. Kupczynski, M. EPR paradox, quantum nonlocality and physical reality. *J. Phys. Conf. Ser*. **2016**, *701*, 012021.
65. Kupczynski, M. On operational approach to entanglement and how to certify it. *Int. J. Quantum Inf*. **2016**, *14*, 1640003.
66. Kupczynski, M. Can we close the Bohr-Einstein quantum debate? *Phil. Trans. R. Soc. A* **2017**, doi:10.1098/rsta.2016.0392.
67. Kupczynski, M. Is Einsteinian no-signalling violated in Bell tests? *Open Phys*. **2017**, *15*, 739–753, doi:10.1515/phys-2017-0087.
68. Kupczynski, M. Quantum mechanics and modeling of physical reality. *Phys. Scr*. **2018**, *93*, 123001, doi:10.1088/1402-4896/aae212.
69. Kupczynski, M. Closing the Door on Quantum Nonlocality. *Entropy* 2018, 20, doi:10.3390/e20110877.
70. Kupczynski, M. Is the Moon there when nobody looks: Bell inequalities and physical reality. *Front. Phys*. **2020**, doi:10.3389/fphy.2020.00273.
71. De Muynck, V.M.; De Baere, W.; Martens, H. Interpretations of quantum mechanics, joint measurement of incompatible observables and counterfactual definiteness. *Found. Phys*. **1994**, *24*, 1589–1664.
72. De Muynck, W.M. *Foundations of Quantum Mechanics;* Kluver Academic: Dordrecht, The Netherlands, 2002.
73. Nieuwenhuizen, T.M. Where Bell went wrong. *AIP Conf. Proc*. **2009**, *1101*, 127–133.
74. Nieuwenhuizen, T.M. Is the contextuality loophole fatal for the derivation of Bell inequalities. *Found. Phys*. **2011**, *41*, 580–591.
75. Pearle, P. Hidden-variable example based upon data rejection. *Phys. Rev. D* **1970**, *2*, 1418–1425.
76. Peres, A. Unperformed experiments have no results. *Am. J. Phys*. **1978**, *46*, 745–7, doi:10.1119/1.11393.
77. Pitovsky, I. George Boole's conditions of possible experience and the quantum puzzle. *Brit. J. Phil. Sci*. **1994**, *45*, 95–125.
78. De la Peña, L.; Cetto, A.M.; Brody, T.A. On hidden variable theories and Bell's inequality. *Lett. Nuovo Cimento* **1972**, *5*, 177.
79. Cetto, A.M.; Valdes-Hernandez, A.; de la Pena, L. On the spin projection operator and the probabilistic meaning of the bipartite correlation function. *Found. Phys*. **2020**, *50*, 27–39.
80. De Raedt, H.; De Raedt, K.; Michielsen, K.; Keimpema, K.; Miyashita, S. Event-based computer simulation model of Aspect-type experiments strictly satisfying Einstein's locality conditions. *J. Phys. Soc. Jap*. **2007**, *76*, 104005.
81. De Raedt, H.; De Raedt, K.; Michielsen, K.; Keimpema, K.; Miyashita, S. Event-by-event simulation of quantum phenomena: Application to Einstein-Podolsky-Rosen-Bohm experiments. *J. Comput. Theor. Nanosci*. **2007**, *4*, 957–991.
82. Zhao, S.; De Raedt, H.; Michielsen, K. Event-by-event simulation model of Einstein-Podolsky-Rosen-Bohm experiments. *Found. Phys*. **2008**, *38*, 322–347.
83. De Raedt, H.; Hess, K.; Michielsen, K. Extended Boole-Bell inequalities applicable to quantum theory. *J. Comp. Theor. Nanosci*. **2011**, *8*, 10119.
84. De Raedt, H.; Michielsen, K.; Jin, F. Einstein-Podolsky-Rosen-Bohm laboratory experiments: Data analysis and simulation. *AIP Conf. Proc*. **2012**, *1424*, 55–66.
85. De Raedt, H.; Jin, F.; Michielsen, K. Data analysis of Einstein-Podolsky-Rosen-Bohm laboratory experiments. *Proc. SPIE* 2013, 8832, N1–N11.
86. De Raedt, H.; Michielsen, K.; Hess, K. The photon identification loophole in EPRB experiments: Computer models with single-wing selection. *Open Phys*. **2017**, *15*, 713–733, doi:10.1515/phys-2017-0085.
87. Thompson, C.H. The chaotic ball: An intuitive analogy for EPR experiments. *Found. Phys. Lett*. **1996**, *9*, 357–382, doi:10.1007/BF02186307.
88. Wigner, E.P. On hidden variables and quantum mechanical probabilities. *Am. J. Phys*. **1970**, *38*, 1005.
89. Żukowski, M.; Brukner, Č. Quantum non-locality—It ain't necessarily so. *J. Phys. A Math. Theor*. **2014**, *47*, 424009.
90. Svozil, K. Quantum hocus-pocus. *ESEP* **2016**, *16*, 25-30, doi:10.3354/esep00171.
91. Khrennikov, A. Bohr against Bell: Complementarity versus nonlocality. *Open Phys*. **2017**, *15*, 734–738.
92. Kracklauer, A.F. Bell's "Theorem": Loopholes vs. conceptual flaws. *Open Phys*. **2017**, 15, doi:10.1515/phys-2017-0088.
93. Boughn, S. Making sense of Bell's theorem and quantum nonlocality. *Found. Phys*. **2017**, *47*, 640–657.
94. Hess, K. Bell's Theorem and instantaneous influences at a distance. *J. Mod. Phys*. **2018**, *9*, 1573-1590, doi:10.4236/jmp.2018.98099.
95. Jung, K. Violation of Bell's inequality: Must the Einstein locality really be abandoned? *J. Phys. Conf. Ser*. **2017**, *880*, 012065.
96. Jung, K. Polarization correlation of entangled photons derived without using non-local interactions. *Front. Phys*. **2020**, doi:10.3389/fphy.2020.00170.
97. Griffiths, R.B. Nonlocality claims are inconsistent with Hilbert-space quantum mechanics. *Phys. Rev A* **2020**, *101*, 022117, doi:10.1103/PhysRevA.101.022117.
98. Basieva, I.; Cervantes, V.H.; Dzhafarov, E.N.; Khrennikov, A. True contextuality beats direct influences in human decision making. *J. Exp. Psychol. Gen*. **2019**, *148*, 1925-1937.
99. Cervantes, V.H.; Dzhafarov, E.N. True contextuality in a psychophysical experiment. *J. Math. Psychol*. **2019**, *91*, 119–127.

100. Aerts, D.; Argüelles, J.; Beltran, L.; Geriente, S.; Sassoli de Bianchi, M.; Sozzo, S.; Veloz, T. Quantum entanglement in physical and cognitive systems: A conceptual analysis and a general representation. *Eur. Phys. J. Plus* **2019**, *134*, 493, doi:10.1140/epjp/i2019-12987-0.
101. Araujo, M.; Quintino, M.T.; Budroni, C.; Cunha, M.T.; Cabello, A. All noncontextuality inequalities for the n-cycle scenario. *Phys. Rev. A* **2013**, *88*, 022118.
102. Boole, G. On the theory of probabilities. *Philos. Trans. R. Soc. Lond.* **1862**, *152*, 225–252.
103. Suppes, P.; Zanotti, M. When are probabilistic explanations possible? *Synthese* **1981**, *48*, 191–199.
104. Leggett, A.J.; Garg, A. Quantum mechanics versus macroscopic realism: Is the flux there when nobody looks. *Phys. Rev. Lett.* **1985**, *9*, 857–860.
105. Clauser, J.F.; Horne, M.A.; Shimony, A.; Holt, R.A. Proposed experiment to test local hidden-variable theories. *Phys. Rev. Lett.* **1969**, *23*, 880.
106. Klyachko, A.A.; Can, M.A.; Binicioglu, S.; Shumovsky, A.S. Simple test for hidden variables in spin-1 systems. *Phys. Rev. Lett.* **2008**, *101*, 020403.
107. Lapkiewicz, R.; Li, P.; Schaeff, C.; Langford, N.K.; Ramelow, S.; Wiesniak, M.; Zeilinger, A. Experimental nonclassicality of an indivisible quantum system. *Nature* **2011**, *474*, 490–493.
108. Bohm, D. *Quantum Theory*; Prentice Hall: New York, NY, USA, 1989; ISBN 0486659690.
109. Bell, J.S. On the Einstein-Podolsky-Rosen paradox. *Physics* **1964**, *1*, 195–200.
110. Adenier, G.; Khrennikov, A. Is the fair sampling assumption supported by EPR experiments? *J. Phys. B Atom. Mol. Opt. Phys.* **2007**, *40*, 131–141.
111. Adenier, G.; Khrennikov, A. Test of the no-signaling principle in the Hensen loophole-free CHSH experiment. *Fortschr. Der Phys.* **2017**, doi:10.1002/prop.201600096.
112. Bednorz, A. Analysis of assumptions of recent tests of local realism. *Phys. Rev. A* **2017**, *95*, 042118.
113. Lin, P.S.; Rosset, D.; Zhang, Y.; Bancal, J.D.; Liang, Y.C. Device-independent point estimation from finite data and its application to device-independent property estimation. *Phys. Rev. A* **2018**, *97*, 032309, doi:10.1103/PhysRevA.97.032309.
114. Zhang, Y.; Glancy, S.; Knill, E. Asymptotically optimal data analysis for rejecting local realism. *Phys. Rev. A* **2011**, *84*, 062118, doi:10.1103/PhysRevA.84.062118.
115. Christensen, B.G.; Liang, Y.-C.; Brunner, N.; Gisin, N.; Kwiat, P. Exploring the limits of quantum nonlocality with entangled photons. *Phys. Rev. X* **2015**, *5*, 041052, doi:10.1002/prop.201600096.
116. Liang, Y.C.; Zhang, Y. Bounding the plausibility of physical theories in a device-independent setting via hypothesis testing. *Entropy* **2019**, *21*, 185, doi:10.3390/e21020185.
117. Kupczynski, M. Significance tests and sample homogeneity loophole. *arXiv* **2015**, arXiv:1505.06349.
118. Kupczynski, M.; De Raedt, H. Breakdown of statistical inference from some random experiments. *Comp. Phys. Commun.* **2016**, *200*, 168.
119. Larsson, J.-A. Loopholes in Bell inequality tests of local realism. *J. Phys. A Math. Theor.* **2014**, *47*, 424003, doi:10.1088/1751-8113/47/42/424003.
120. Myrvold, W.; Genovese, M., Shimony, A. "Bell's Theorem". In *The Stanford Encyclopedia of Philosophy*, Fall 2020 ed.; Zalta, E.N., Ed. Available online: https://plato.stanford.edu/archives/fall2020/entries/bell-theorem/ (accessed on 17 May 2021).
121. Blasiak, P.; Pothos, E.M.; Yearsley, J.M.; Gallus, C.; Borsuk, E. Violations of locality and free choice are equivalent resources in Bell experiments. *Proc. Natl. Acad. Sci. USA* **2021**, *118*, doi:10.1073/pnas.2020569118.
122. Bell, J.S. The theory of local beables. *Epistemol. Lett.* **1976**, *9*, 11–24.
123. Bell, J.S. *Speakable and Unspeakable in Quantum Mechanics*, 2nd ed.; Cambridge University Press: Cambridge, UK, 2004.
124. Kupczynski, M. A comment on: The violations of locality and free choice are equivalent resources in Bell experiments. *arXiv* **2021**, arXiv:2105.14279.
125. Kofler, J.; Giustina, M.; Larsson, J.-Å.; Mitchel, M.W. Requirements for loophole-free photonic Bell test using imperfect generators. *Phys. Rev. A* **2016**, *93*, 032115.
126. Svozil, K. Quantum violation of the Suppes-Zanotti inequalities and "contextuality". *Int. J. Theor. Phys.* **2021**, *60*, 2300–2310.
127. Calude, C.S.; Dinneen, M.J.; Dumitrescu, M.; Svozil, K. Experimental evidence of quantum randomness incomputability. *Phys. Rev. A* **2010**, *82*, 022102.
128. Martínez, A.C.; Solis, A.; Rojas, R.D.H.; U'Ren, A.B.; Hirsch, J.G.; Castillo, I.P. Advanced statistical testing of quantum random number generators. *Entropy* **2018**, *20*, 886, doi:10.3390/e20110886.
129. Iannuzzi, M.; Francini, R.; Messi, R.; Moricciani, D. Bell-type polarization experiment with pairs of uncorrelated optical photons. *Phys. Lett. A* **2020**, *384*, 126200, doi:10.1016/j.physleta.2019.126200.
130. Raussendorf, R. Contextuality in measurement-based quantum computation. *Phys. Rev. A* **2013**, *88*, 022322
131. Howard, M.; Wallman, J.; Veitch, V.; Emerson, J. Contextuality supplies the 'magic' for quantum computation. *Nat. Cell Biol.* **2014**, *510*, 351–355, doi:10.1038/nature13460
132. Abramsky, S.; Barbosa, R.S.; Mansfield, S. Contextual fraction as a measure of contextuality. *Phys. Rev. Lett.* **2017**, *119*, 050504, doi:10.1103/PhysRevLett.119.050504.
133. Abramsky, S.; Brandenburger, A. The sheaf-theoretic structure of non-locality and contextuality. *N. J. Phys.* **2011**, *13*, 113036.
134. Cabello, A.; Severini, S.; Winter, A. Graph-Theoretic Approach to Quantum Correlations. *Phys. Rev. Lett.* **2014**, *112*, 040401, doi:10.1103/physrevlett.112.040401.

135. Acín, A.; Fritz, T.; Leverrier, A.; Sainz, A.B. A combinatorial approach to nonlocality and contextuality. *Commun. Math. Phys.* **2015**, *334*, 533–628.
136. Spekkens, R.W. Contextuality for preparations, transformations, and unsharp measurements. *Phys. Rev. A* **2005**, *71*, 052108, doi:10.1103/PhysRevA.71.052108.